\documentclass{article}
\usepackage{spconf,amsmath,graphicx}

\usepackage{url,amsfonts,amssymb,bm}

\usepackage{enumitem}
\usepackage{algorithm}
\usepackage{algorithmicx, algpseudocode }
\usepackage{multirow}
\usepackage{color}
\usepackage{mathtools}
\usepackage{caption}
\usepackage{subcaption}
\usepackage{amsthm}

\usepackage[colorlinks=true, linkcolor=blue, citecolor=blue, urlcolor=blue]{hyperref}
\hypersetup{pdfcreator  = {LaTeX with hyperref package},
               pdfproducer = {dvips + ps2pdf} }

\usepackage{tikz}

\usetikzlibrary{calc}

\usepackage{tikz,spconf,amsmath,graphicx,bbm}
\usepackage{mathrsfs}
\usepackage{bbm}

\usetikzlibrary{positioning}

\newtheorem{theorem}{Theorem}

\theoremstyle{definition}

\DeclareMathOperator*{\mmse}{MMSEE}

\DeclareMathOperator*{\var}{var}
\DeclareMathOperator*{\E}{\mathbb{E}}

\usepackage{bm}

\long\def\comment#1{}

\newfont{\bbb}{msbm10 scaled 700}

\newfont{\bb}{msbm10 scaled 1100}

\newcommand{\uv}{{\bf u}}

\newcommand{\xv}{{\bf x}}
\newcommand{\yv}{{\bf y}}
\newcommand{\zv}{{\bf z}}

\newcommand{\Dc}{{\cal D}}

\newcommand{\Nc}{{\cal N}}

\title{  Compression of user generated content using denoised references}
\name{Eduardo Pavez$^\star$, Enrique Perez$^\star$, Xin Xiong$^\star$, Antonio Ortega$^\star$, Balu Adsumilli$^\dagger$ \thanks{This work was funded in part by a gift from YouTube. Author's email: \{pavezcar, perezenr, xiongxin, aortega\}@usc.edu  }}
\address{$^\star$University of Southern California, Los Angeles, California, USA 
\\$^\dagger$Google Inc, Mountain View, CA }
\begin{document}
% Uncomment to reduce the margins above and below math environments
% \setlength{\abovedisplayskip}{4pt}
% \setlength{\belowdisplayskip}{4pt}
\ninept
\maketitle
\begin{abstract}
Video shared over the internet is commonly referred to as user generated content (UGC). UGC video  may have low quality due to  various factors including previous compression.  UGC video is uploaded by users, and then it is re-encoded to be made available at various levels of quality. In a traditional video coding pipeline  the encoder parameters are  optimized to minimize a rate-distortion criterion,   but when the input signal has low quality, this results in sub-optimal coding parameters optimized to preserve undesirable artifacts. In this paper we formulate the UGC compression problem as that of compression of a noisy/corrupted source. The noisy source coding theorem reveals that an optimal UGC compression system is comprised of  optimal denoising of the UGC signal, followed by compression of the denoised signal. Since optimal denoising is unattainable and users may be against modification of their content, we propose encoding the UGC signal, and using denoised references only to compute distortion,  so the encoding process can be guided towards  perceptually better solutions. We demonstrate the effectiveness of the proposed strategy for JPEG compression of UGC images and videos. 
\end{abstract}
\section{Introduction}
Video sharing applications (e.g., YouTube, TikTok) produce a large percentage of Internet traffic. This type of video is commonly referred to as user generated content (UGC) \cite{wang2019youtube}.  UGC is first uploaded by users and then it is  re-encoded by service providers in order to be made available at various levels of quality and resolution. The traditional video compression pipeline assumes the input video is pristine, however this is often not true for UGC, where the source material has been compressed by the users sharing it. In addition, UGC   may have low  quality due to additional factors, e.g., use of non professional video equipment, poor shooting skills, low light,   editing,   special effects, etc. 

In the traditional video compression pipeline, the encoder-decoder parameters are optimized to minimize the distortion subject to   bitrate (and   computational complexity) constraints \cite{ortega1998rate,sullivan1998rate}. However, when   distortion is computed with respect to a corrupted reference signal, the rate distortion optimization process may lead to sub-optimal coding parameters that preserve  undesirable features that do not improve perceptual quality (e.g., blocking artifacts due to previous compression).   
The fundamental  UGC compression problem is to,   
\emph{given an UGC signal and a compression system (e.g., JPEG, AV1), choose coding parameters to accurately represent the perceptually meaningful parts of the signal, while avoiding allocating resources to encode compression artifacts and noise. }

To address the issue of a low quality and unreliable reference,  researchers have proposed using non reference metrics to assess subjective video quality \cite{yu2021predicting,mittal2012no}, which can be used to  perceptually optimize (guide) the compression of UGC videos. 
Another approach classifies UGC based on content category and   similarity in rate-distortion characteristics, so that fixed  coding parameters can be used for each UGC class \cite{john2020rate,ling2020towards}. Because of the negative effect of   noise,  the importance of   denoisers as part of the UGC  coding pipeline has also been studied \cite{hanooman2021effect}.
While previous works have recognized that the encoding process should adapt to the quality of the input UGC video, and have provided tools and  insights   to  design  UGC compression systems,   we take a step towards solving the UGC compression problem   from a rate-distortion theoretic perspective \cite{berger1971rate}. 
%Source coding theory has helped understand the fundamental limitations of compression systems in terms of rate-distortion trade-offs \cite{berger1971rate}, while rate-distortion optimization techniques are at the core of modern video codecs, and provide tools to choose optimal coding parameters \cite{ortega1998rate,sullivan1998rate}. 
% By using an  appropriate rate-distortion (RD) formulation,  we   reveal that optimal denoising/estimation is important. 
% We then propose a practical application of this result, that uses denoised references to guide  parameter selection  of a traditional  encoder. 
%
\begin{figure}[t]
    \centering
    \includegraphics[width=0.35\textwidth]{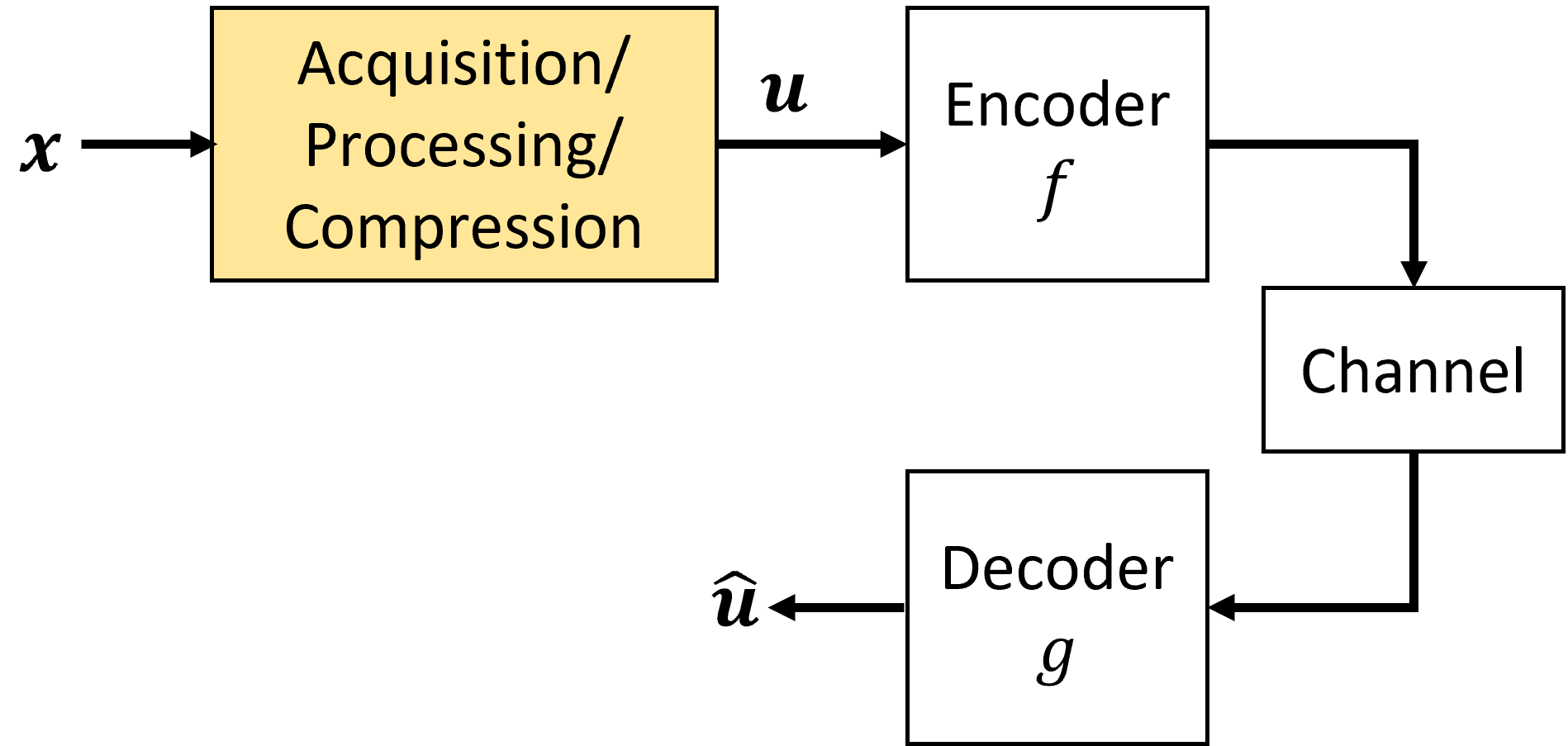}
    \caption{Block diagram for UGC video coding. $\xv$ is the pristine (unknown) signal, $\uv$ is the UGC video, and $\hat{\uv}$ is the decoded signal. }
    \label{fig:UGC_coding}
\end{figure}
In \autoref{sec_ugc_formulation} we formulate the UGC compression problem  as an instance of noisy source coding, where the noiseless source corresponds to the pristine original, and the noisy/corrupted signal is the UGC. This process is depicted in \autoref{fig:UGC_coding}. 
In this ideal scenario, the goal is to minimize distortion  computed with respect to the pristine (unknown) original. By invoking a noisy source coding theorem, we can show that the optimal encoder-decoder system in the mean-squared-error (MSE) sense is comprised of optimal MSE estimation of the clean source from the noisy source (i.e., optimal denoising), followed by optimal (noiseless) source coding of this estimate \cite{dobru1962source,wolf1970transmission,berger1971rate}. The noisy source coding theorem has been applied to compression of noisy images \cite{al1998lossy}, speech coding \cite{ephraim1988unified,fischer1990estimation,gibson1991filtering} and to the design  of video coding systems robust to pre- and post-processing \cite{dar2016postprocessing,dar2018optimized}. However, to the best of our knowledge, it has not been used  yet to study UGC compression.

\begin{figure}[t]
    \centering
\includegraphics[width=0.4\textwidth]{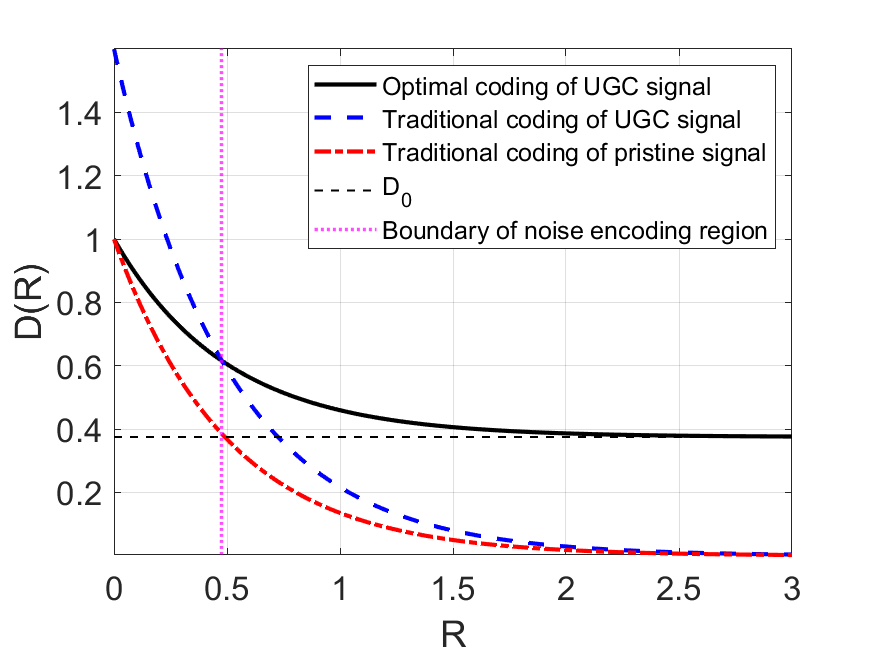}
    \caption{ Optimal distortion-rate functions for scalar zero mean Gaussian sources. Comparison of optimal RD curves for encoding: $x$ with reference $x$, the noisy signal $u=x+\eta$ with $u$ as reference ($D(R)$ from \eqref{eq_RD_traditional}), and $u$ with pristine source $x$ as reference  ($D^{\text{ugc}}(R)$ from \eqref{eq_RD_ugc}).  $x$ has variance $\sigma_x^2 = 1$, the UGC signal is  $u = x + \eta$, where $\eta$ is a zero mean Gaussian with variance $\sigma_{\eta}^2 = 0.6$, independent of $\xv$. }
    \label{fig:optimal_RD_gaussian}
\end{figure}

In traditional video coding the distortion goes to zero as the rate increases and the quality of the encoded video improves. In contrast, a consequence of the noisy source coding theorem is that distortion   should not go to zero, i.e., further increases in bitrate beyond a certain point do not lead to improved performance. This is because in the UGC coding scenario,  we wish to minimize distortion with respect to the pristine reference, but we have to do this without being able to encode the pristine signal directly.   
An example of the optimal distortion-rate curve for a Gaussian source corrupted by additive Gaussian noise \cite{wolf1970transmission} is depicted in \autoref{fig:optimal_RD_gaussian}. When distortion is computed with respect to the noisy input, the distortion quickly goes to zero, while for the optimal UGC coding system, distortion decreases more slowly and saturates  to a positive distortion value.

The theoretically optimal UGC compression system  uses an optimally denoised UGC signal both as a reference (for distortion computation) and as a source (for encoding). 
However, such system cannot be implemented  in practice because we cannot guarantee that an optimal denoiser can be found. As a practical alternative we propose  using off-the-shelf denoisers to compute a denoised UGC signal to be used for distortion computation as a replacement for the (unavailable) pristine original, while  the UGC signal  is used as a source for the encoder.
Although a system that encodes a denoised UGC signal (using a sub-optimal denoiser) could be used, we  propose encoding the UGC signal directly instead, because:
1)  the denoiser  we use is no longer the optimal one, thus  a system that encodes a sub-optimally denoised UGC signal is also sub-optimal,
2) finding good denoising/restoration algorithms may be difficult,  given that there may be multiple reasons for quality degradation in a UGC signal and thus a specific denoiser may  not always produce  reliable outputs,
3) the wrong choice of denoiser parameters may result into  encoding a lower quality source than the UGC input, and
4) users may  object to a service provider modifying their uploaded content.
%Using a denoised reference allows us to determine when we have reached the maximum desirable bitrate for encoding the UGC input. However, as the bitrate increases, the encoded signal approaches the UGC signal and not the denoised reference, but because   the distortion-rate curve saturates at a lower bitrate, we can avoid bitrates that encode noise and artifacts.

%While encoding a denoised signal is theoretically optimal, in a  practical system it may be preferable to encode the UGC signal directly instead, because: 
%1) users may  object to a service provider modifying their uploaded content, and 
%2) finding good denoising/restoration algorithms may be difficult, given that there may be multiple reasons for quality degradation in a UGC signal and thus a specific denoiser may  not always produce  reliable outputs. 
%Hence, we propose to use a denoised UGC signal only as a proxy to compute distortion, i.e., as a replacement for the (unavailable) pristine original, while using the UGC signal itself as the source for the encoder. 
In \autoref{sec_exp} we show experimentally that the rate-distortion curve of the proposed system has a saturation region, similar to the  optimal RD curve. 
We use this and propose an algorithm to choose coding parameters based on detecting saturation of the distortion curve, to avoid encoding at bitrates for which the encoded signal quality does not improve. We show that for a JPEG encoder, the quality parameter associated  with the onset of the saturation region is positively correlated with the perceptual quality of the UGC video. 
\section{The UGC compression problem}
\label{sec_ugc_formulation}
In this section we propose a theoretical formulation of the UGC compression problem. We show that optimal denoising is essential for efficient compression of UGC. We then propose a practical framework using an off-the-shelf denoiser. %
\subsection{Noisy source coding}
In  UGC compression  (\autoref{fig:UGC_coding}),    $\xv$ and $\uv$ are random vectors  representing the  pristine content and the UGC signal, respectively.
The encoded representation is denoted by $b = f(\uv)$, where $f$ is the encoder, and  the bit rate of this representation  is denoted by $\ell(b)$. The output of the decoder is 
denoted by  $\hat{\uv} = g(f(\uv))$, where $g$ is the decoder. A traditional (noiseless) source coding problem has the form 
\begin{equation}\label{eq_RD_traditional}
  D(R) =   \min_{f,g} \E[\Vert \uv - \hat{\uv} \Vert^2] \textnormal{ s.t. } \ell(f(\uv)) \leq R,
\end{equation}
where $R$ is the rate, and $D(R)$ is the distortion-rate function.  Note that as the bitrate $R$ increases in \eqref{eq_RD_traditional}, $\hat{\uv}\rightarrow \uv$  and   the distortion $D(R)$ decreases so that  $\lim_{R \rightarrow \infty}D(R) = 0$. This is problematic when the input $\uv$ is UGC,  because at high rates, low distortion simply means that $\uv$ and $\hat{\uv}$ are close, but the best possible representation ($\uv$, corresponding to $D(R)=0$) may not  have good quality.

Ideally, since the source is noisy,  the source coding problem should be formulated so that  distortion is computed with respect to the pristine original:
 \begin{equation}
  D^{\text{ugc}}(R) =   \min_{f,g} \E[\Vert \xv - \hat{\uv} \Vert^2] \textnormal{ s.t. } \ell(f(\uv)) \leq R.
  \label{eq:opt-pristine}
\end{equation}
Under the optimality criterion  of \eqref{eq:opt-pristine},     
%rate distortion function for UGC coding 
%$D^{\text{ugc}}(R)$ 
the decoded signal $\hat{\uv}$ has to approximate the pristine content $\xv$. The  following result allows us to break down \eqref{eq:opt-pristine} into two simpler steps: 1) optimal denoising, and 2) optimal (noiseless) source coding.
\begin{theorem}\cite{dobru1962source,wolf1970transmission,berger1971rate}\label{th_noisy_source_coding}
The optimal distortion-rate function for the UGC coding problem is: 
\begin{equation}\label{eq_RD_ugc}
  D^{\text{ugc}}(R) = D_0 + \min_{\bar{f},\bar{g}} \E[\Vert \yv - \hat{\uv} \Vert^2] \textnormal{ s.t. } \ell(\bar{f}(\yv)) \leq R,
\end{equation}
\text{where} $D_0 = \E[\Vert \xv - \yv \Vert^2]$,  $\yv = \E[\xv \mid \uv]$, \text{and} $\hat{\uv} = \bar{g}(\bar{f}(\yv))$.
\end{theorem}
Note that $\yv$ is the minimum mean square error estimator (MMSEE) of the pristine signal, which does not depend on the encoder-decoder functions, or the rate $R$. 
The proof of \autoref{th_noisy_source_coding} \cite{dobru1962source,wolf1970transmission,berger1971rate} uses two facts: (i)  $\yv$ and  $(f \circ g)(\yv)$ are measurable functions of $\uv$, and (ii) the MMSEE is orthogonal, namely $\E[(\xv - \yv)h(\uv)]=0$, for any measurable function $h$.
A first consequence of \autoref{th_noisy_source_coding} is the  lower bound 
\begin{equation}\label{eq_lower_bound_ideal_ugc}
    D^{\text{ugc}}(R) \geq   D_0, 
\end{equation}
%Property \eqref{eq_lower_bound_ideal_ugc} 
which establishes that $D_0$ is the lowest achievable distortion by any encoder-decoder at any rate. 
Since $D_0$ is the error of the MMSEE, it  can be interpreted as a  quality metric of the UGC signal.
Another consequence of \autoref{th_noisy_source_coding} is that for  an encoder-decoder to  asymptotically achieve this lower bound, that is,  $\lim_{R \rightarrow \infty}D^{\text{ugc}}(R) = D_0$, we have that $f = \bar{f} \circ \mmse$, while  $g = \bar{g}$ is the corresponding decoder for $\bar{f}$.  In other words, an optimal UGC compression system has two components: 1) optimal  denoising with the MMSEE, $\yv = \E[\xv \mid \uv]$, and 2) optimal lossy  encoding-decoding that acts on $\yv$ instead of $\uv$.
%\begin{enumerate}
%    \item Optimal  denoising with the MMSEE, $\yv = \E[\xv \mid \uv]$, 
%    \item Lossy encoding/decoding that acts on $\yv$ instead of $\uv$.  
%\end{enumerate}
In \autoref{fig:optimal_RD_gaussian} we depict   optimal distortion-rate curves for  a zero mean scalar Gaussian source $x$, contaminated with additive independent zero mean Gaussian noise $\eta$. We can observe that the optimal UGC compression system saturates at distortion $D_0$ when  $R$ is large, while traditional source coding systems converge faster to zero distortion.    For this example, we can also compare the derivatives of $D(R)$ from \eqref{eq_RD_traditional} and $D^{\textnormal{ugc}}(R)$ from \eqref{eq_RD_ugc}. Applying closed form expressions for $D(R)$ and  $D^{\textnormal{ugc}}(R)$ from  \cite{wolf1970transmission} we  obtain
\begin{equation}
    \left\vert \frac{\partial D^{\text{ugc}}(R)}{\partial R}\right\vert = \left(\frac{\var(x)}{\var(u)}\right)^2 \left\vert \frac{\partial D(R)}{\partial R} \right\vert \leq \left\vert  \frac{\partial D(R)}{\partial R} \right\vert,
\end{equation}
where  $\var(x)/\var(u) = \var(x)/(\var(x) + \var(\eta)) \leq 1$. Therefore,  the traditional formulation (i.e., \eqref{eq_RD_traditional}), suggests  that by  increasing the rate by $1$ bit,  quality improves by $\left \vert {\partial D(R)}/{\partial R} \right\vert$, while in fact a  correct formulation (i.e., \eqref{eq_RD_ugc}) only guarantees  the more modest improvement by $\vert {\partial D^{\textnormal{ugc}}(R)}/{\partial R}  \vert$. 
%Therefore for worse quality input UGC signals  (i.e., such that $\var(x)/\var(u)$ is smaller), it becomes more difficult for the decoded UGC signal to the decoded quality achievable if we were using the pristine source. 

While \autoref{th_noisy_source_coding} gives us a clear solution, its implementation is impractical for several reasons. First, optimal MSE denoising depends on the  signal $\xv$, and on the joint distribution of $\xv$ and $\uv$, which are unknown. 
Second, while practical  video codecs can achieve impressive compression performance using rate-distortion 
optimization of encoding parameters  for a single video \cite{ortega1998rate,sullivan1998rate}, 
the noisy  source coding formulation \eqref{eq_RD_ugc} is concerned with guaranteeing optimality in an average sense, i.e., when considering the average performance for a family of  signals $\uv,\xv$ with the same distribution.  
Third, while \autoref{th_noisy_source_coding} suggests directly encoding a denoised signal $\yv$, this may be undesirable for the reasons mentioned in the introduction. 
%: 1) users may be against a service provider modifying their uploaded content, and 2) finding good denoising/restoration algorithms may be difficult, given the varied reasons for low  quality of the UGC signal, and in some cases may produce  unreliable outputs. 
%
%
%
\begin{figure}[t]
    \centering
    \includegraphics[width=0.35\textwidth]{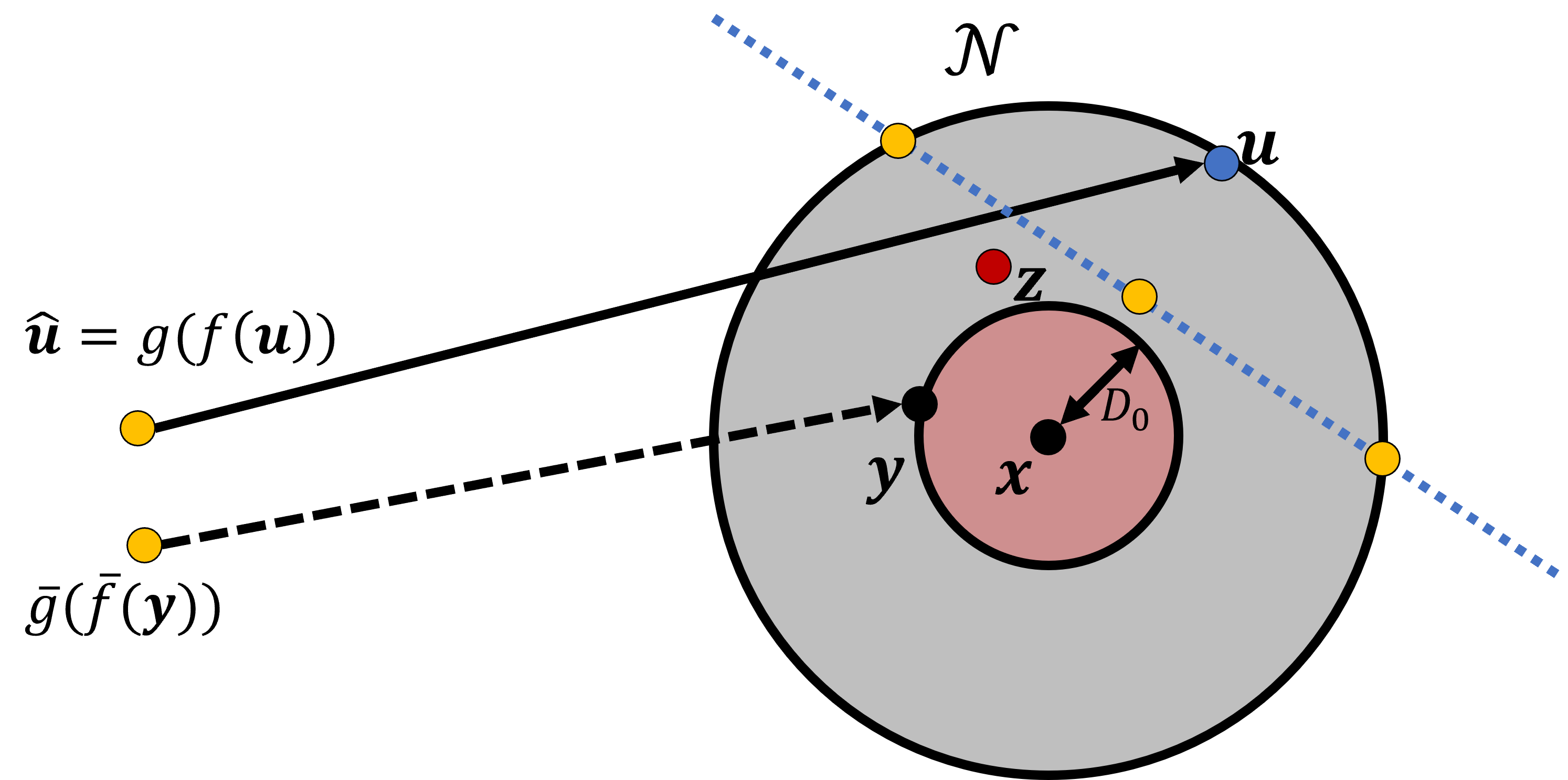}
    \caption{UGC encoding strategies using difference signals for encoder input and  reference metric. }
    \label{fig:rd_optimization_alternative_ref}
\end{figure}
\subsection{UGC compression with denoised references}
We propose compressing the UGC signal $\uv$ using the  signal $\zv= \Dc(\uv)$ as a reference for distortion computation,  where $\Dc(\cdot)$ is a denoiser. Using this metric we can guide the encoding process towards solutions with fewer artifacts. The main idea behind our proposal is illustrated in  \autoref{fig:rd_optimization_alternative_ref}. 
According to the noise source coding theory, points inside the red circle with $\xv$ at the center will have distortions that are not achievable. The MMSEE $\yv$ is a point in the boundary of that red circle and thus ideal UGC compression using $\xv$ as reference is depicted by the dashed arrow, where as the bitrate increases, the encoded signal $\bar{g}(\bar{f}(\yv))$ approaches $\yv$. 
A standard UGC encoder is represented by a solid black arrow, so that as the bitrate increases, the encoded signal $\hat{\uv} = g(f(\uv))$ approaches $\uv$. 
Note that when the bitrate $R$ is small, $\hat{\uv}$  is at similar distance from both the UGC signal and the pristine original, that is  $\Vert \hat{\uv} - \xv\Vert \approx \Vert \hat{\uv} - \uv \Vert \gg \Vert \uv - \xv \Vert$, thus in this regime, an encoded version of $\uv$ may look similar to an encoded version of $\xv$. In this figure we can see that as the rate increases $\hat{\uv}$ will become closer to $\uv$ at the expense of increasing the distance to $\xv$, which would be clearly undesirable. Thus, we can define a \emph{noise encoding region}, corresponding to the family of all encoder-decoders (with their parameters), for which the encoded signal $\hat{\uv}$  is closer to $\uv$ than $\xv$, or more precisely 
\begin{equation}
    \Nc = \lbrace (f,g): \Vert g(f(\uv)) - \xv\Vert > \Vert g(f(\uv)) - \uv \Vert \rbrace.
\end{equation}
The boundary of the noise encoding region is depicted by a blue dashed line in \autoref{fig:rd_optimization_alternative_ref}. For the  optimal UGC compression system from \autoref{fig:optimal_RD_gaussian}, the noise encoding region corresponds to all the $(R,D(R))$ points for which the distortion of the traditional coding system (dashed blue curve) is below the distortion of the optimal UGC coding system (solid black line). %The boundary of the noise encoding region is depicted by a dashed magenta line. 
Our goal is to avoid $\Nc$, and  find encoder-decoder pairs, $(f,g) \notin \Nc$. 
 Clearly, $\Nc$ cannot be found, given that we do not have access to  $\xv$. 
If the denoised signal is a better approximation to the pristine original than the UGC signal, that is,  $\Vert \zv - \xv \Vert < \Vert \uv -\xv \Vert$, then we can use $\zv$ to guide $\hat{\uv}$ away from $\uv$ as the rate $R$ increases.
Thus, we  define an \emph{empirical noise encoding region} using the denoised reference, 
\begin{equation}\label{eq_empirical_noise_encoding}
 \Nc_{\Dc} = \lbrace (f,g): \Vert g(f(\uv)) - \Dc(\uv) \Vert > \Vert g(f(\uv)) - \uv \Vert \rbrace,
\end{equation}
which can be used to  choose coding parameters.
\begin{figure}[t]
    \centering
    \includegraphics[width=0.3\textwidth]{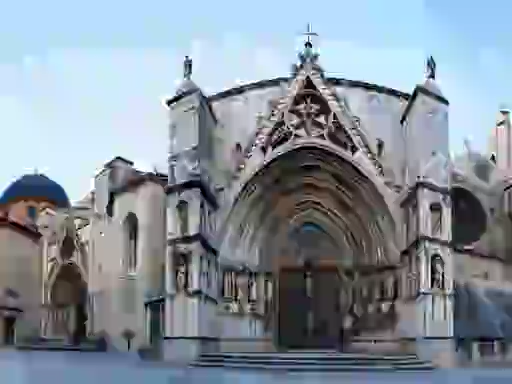}
    \includegraphics[width=0.225\textwidth]{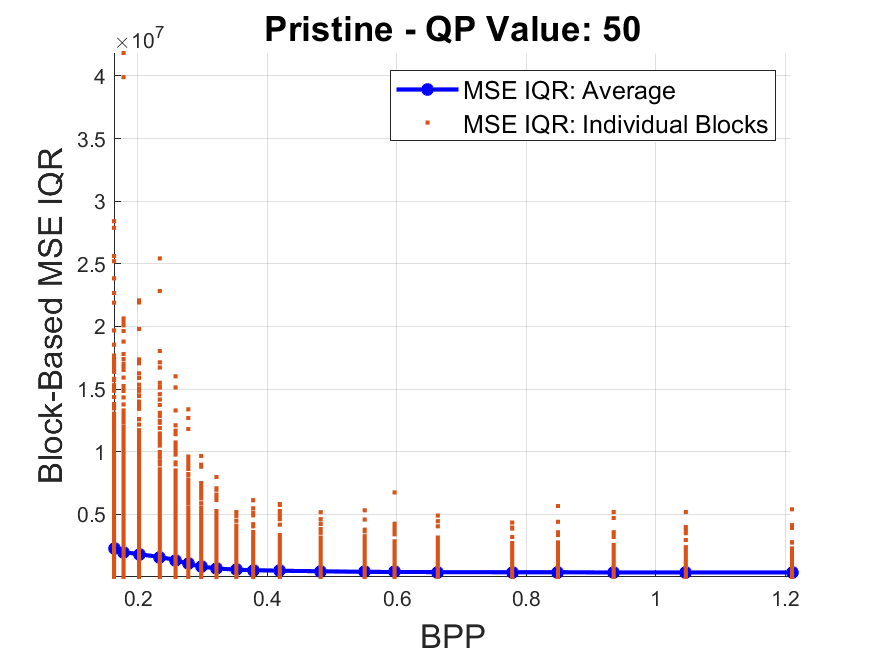}
    \includegraphics[width=0.225\textwidth]{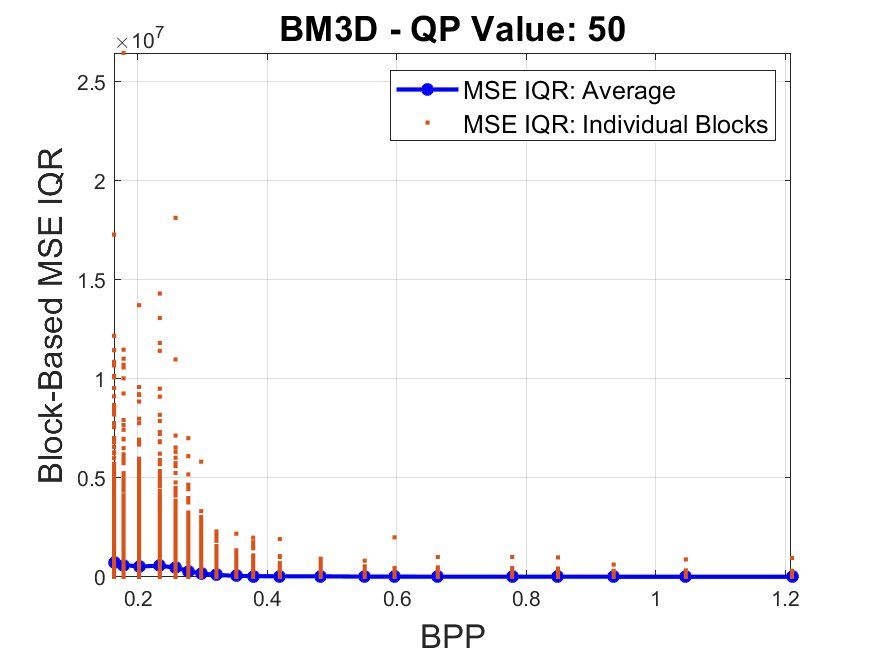}
    \caption{MSE values for individual blocks and their IQR. Synthetic UGC image created by compression with H.264 using quantization parameter $QP=45$ (top). Block MSE with respect to pristine original (left), and block MSE with respect to BM3D denoised reference (right).  }
    \label{fig:blockMSE_rate_saturation}
\end{figure}
\section{Experiments}
\label{sec_exp}
To find the  distortion saturation region, where the quality of the UGC signal does not improve (see \autoref{fig:optimal_RD_gaussian}), we use denoised reference signals and the criteria to detect $\Nc_{\Dc}$ from \eqref{eq_empirical_noise_encoding}. In our  experiments we use JPEG for UGC compression. 
Within a single image,  different regions have varying levels of complexity and thus require different bitrates to achieve the same quality. 
%for example regions with sharp edges and textures, often require higher bitrates for accurate compression. 
Since each  image will have varying mixes of high and low complexity blocks,  we detect saturation of the distortion function at a  $k \times k$ block level. 
Specifically, to capture  the typical block behavior  and to remove outliers, we use the Interquartile Range (IQR) of the per-block MSE. Let $E_i$ be the MSE from the $i$th block (for either pristine reference or denoised reference), and assuming that the $E_i$ are sorted by magnitude, so that $E_i \leq E_{i+1}$ for all blocks, the MSE IQR is defined as $\textnormal{MSE-IQR} = E_{i_2} - E_{i_1}$,
% \begin{equation}
%     \textnormal{MSE-IQR} = E_{i_2} - E_{i_1},
% \end{equation}
 where $E_{i_1}$ is larger than $25\%$ of the values, and $E_{i_2}$ is larger than $75\%$ of the values. The MSE-IQR captures the variation of the middle $50\%$ of the block MSE values, while removing outliers. 
\subsection{Experiments with synthetic UGC images}
\begin{figure}[t]
    \centering
    \includegraphics[width=0.225\textwidth]{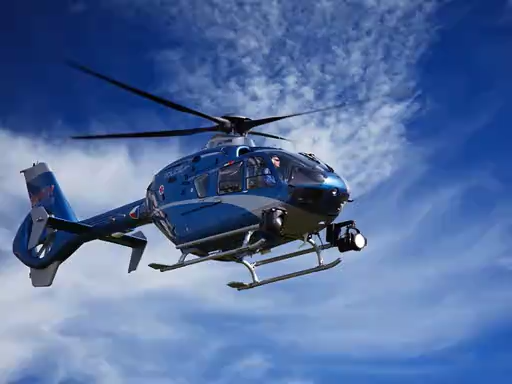}
    \includegraphics[width=0.225\textwidth]{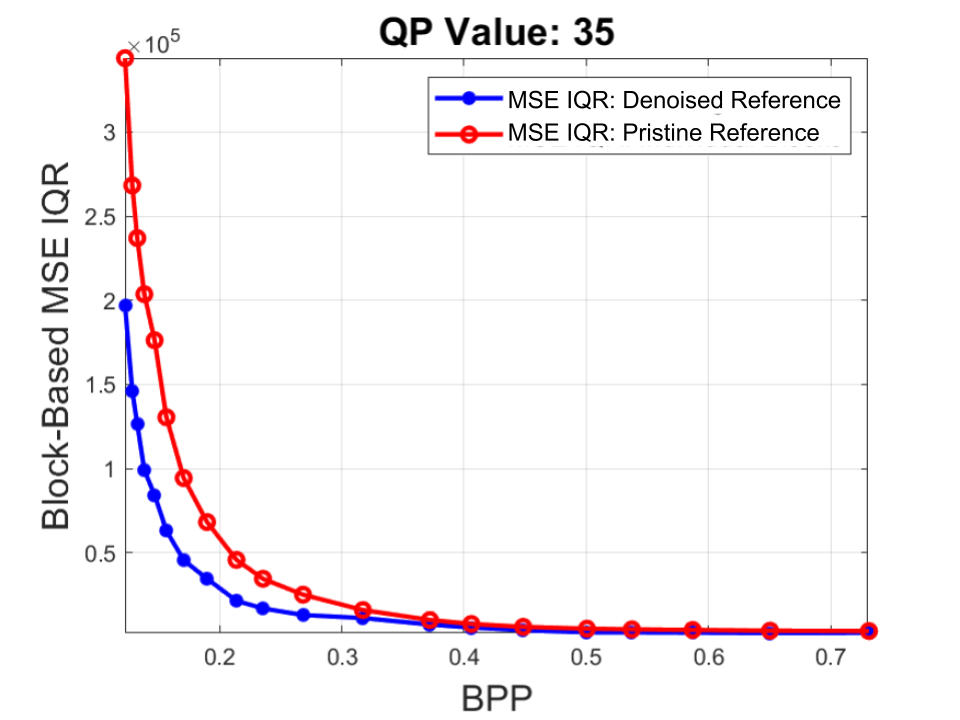}
    \includegraphics[width=0.225\textwidth]{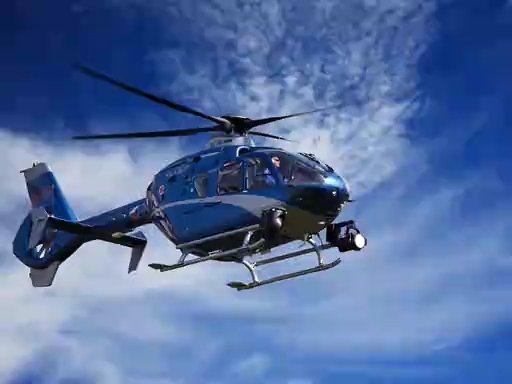}
    \includegraphics[width=0.225\textwidth]{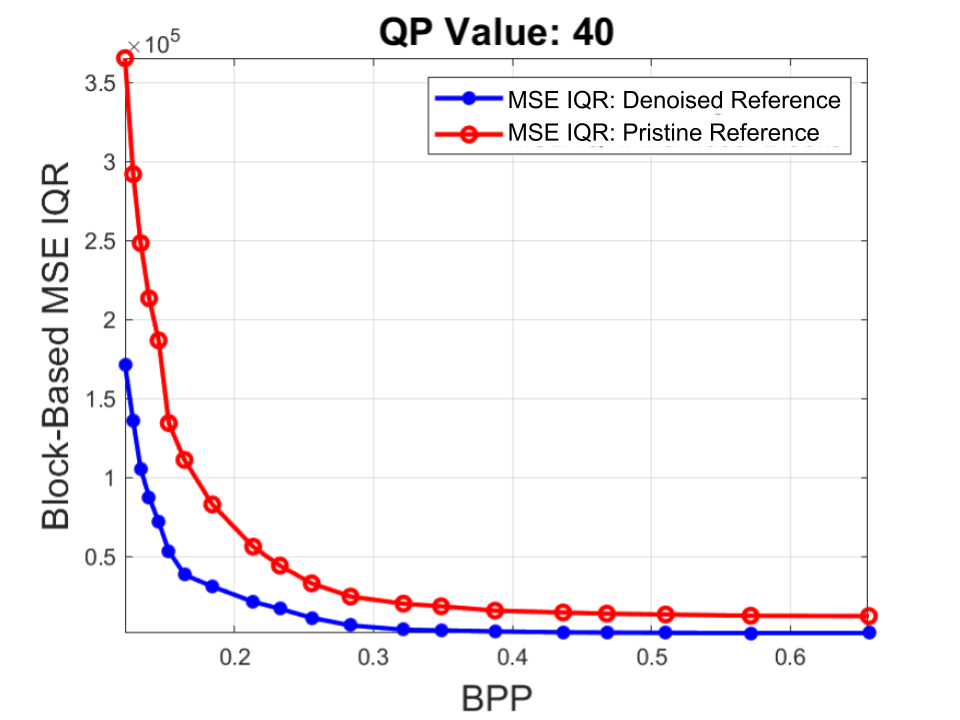}
    \includegraphics[width=0.225\textwidth]{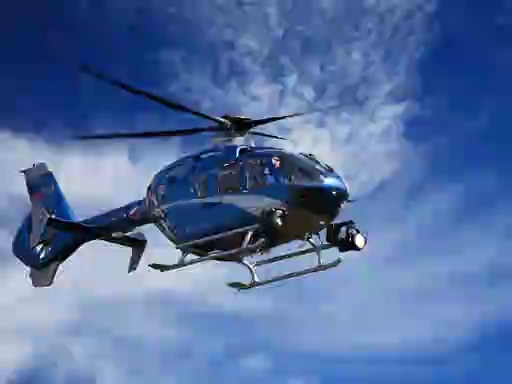}
    \includegraphics[width=0.225\textwidth]{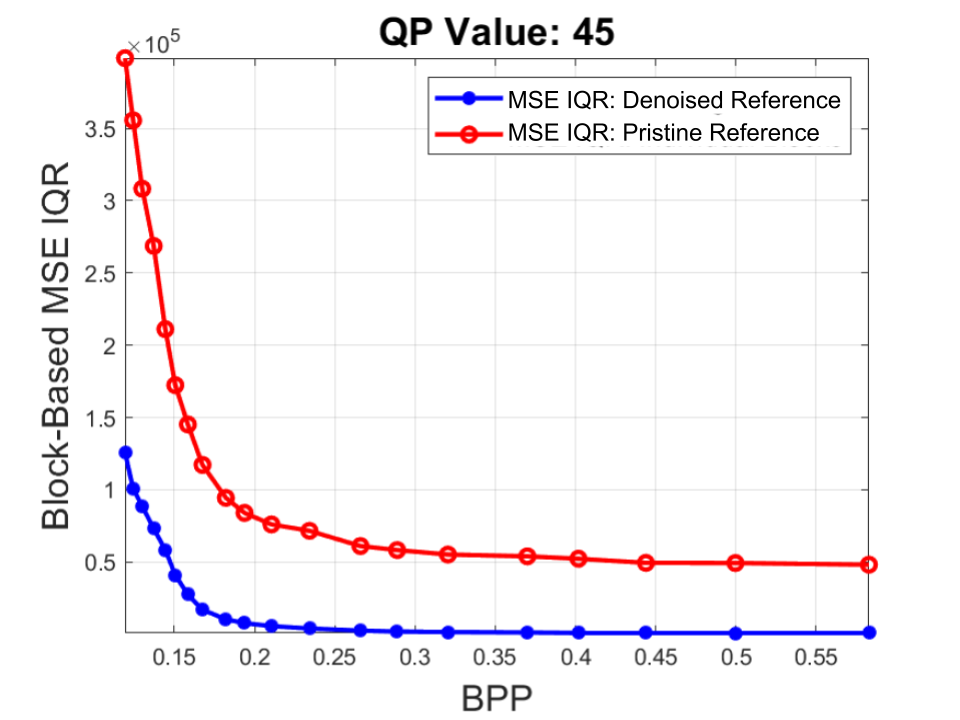}
    \caption{Synthetic UGC images obtained by H.264 compression with high quality (top-left $QP=35$), intermediate quality (middle-left, $QP=40$), and low quality (bottom-left, $QP=45$). Saturation of  block-based MSE-IQR  curves  (right). }
    \label{fig:block_Based_MSEIRQ}
\end{figure}

In this section we show experimentally that when the UGC has low quality  due to previous compression, RD curves computed using the pristine original and the denoised UGC content  have a similar saturation region. We used pristine images from the KADID-10k dataset \cite{kadid10k}, and compressed  them with H.264. In \autoref{fig:blockMSE_rate_saturation}, we depict an example of a heavily compressed image to be used as UGC.  
 This UGC image is then encoded  with JPEG  at $20$ different bitrates. For each bitrate, the image is divided into   $8 \times 8$ blocks, and for each block we compute MSE with respect to the pristine original and 
 with respect to a  \emph{BM3D} denoised \cite{makinen2020collaborative} reference.
 \autoref{fig:blockMSE_rate_saturation} also shows the per block MSE as a function of the total bitrate. At lower bitrate, there is high variation of MSE across blocks, while for higher bitrates, this variation decreases. 
In \autoref{fig:block_Based_MSEIRQ} we plot the MSE-IQR computed for  pristine and denoised references,  as a function of the bitrate.   
For the same UGC image, with different levels of quality, we observe that  both distortions, based on pristine and alternative references,  saturate at similar bitrates, which decrease with the UGC quality level. 
\subsection{Experiments with YouTube UGC dataset}
\begin{figure}[t]
    \centering
    \includegraphics[width=0.47\textwidth]{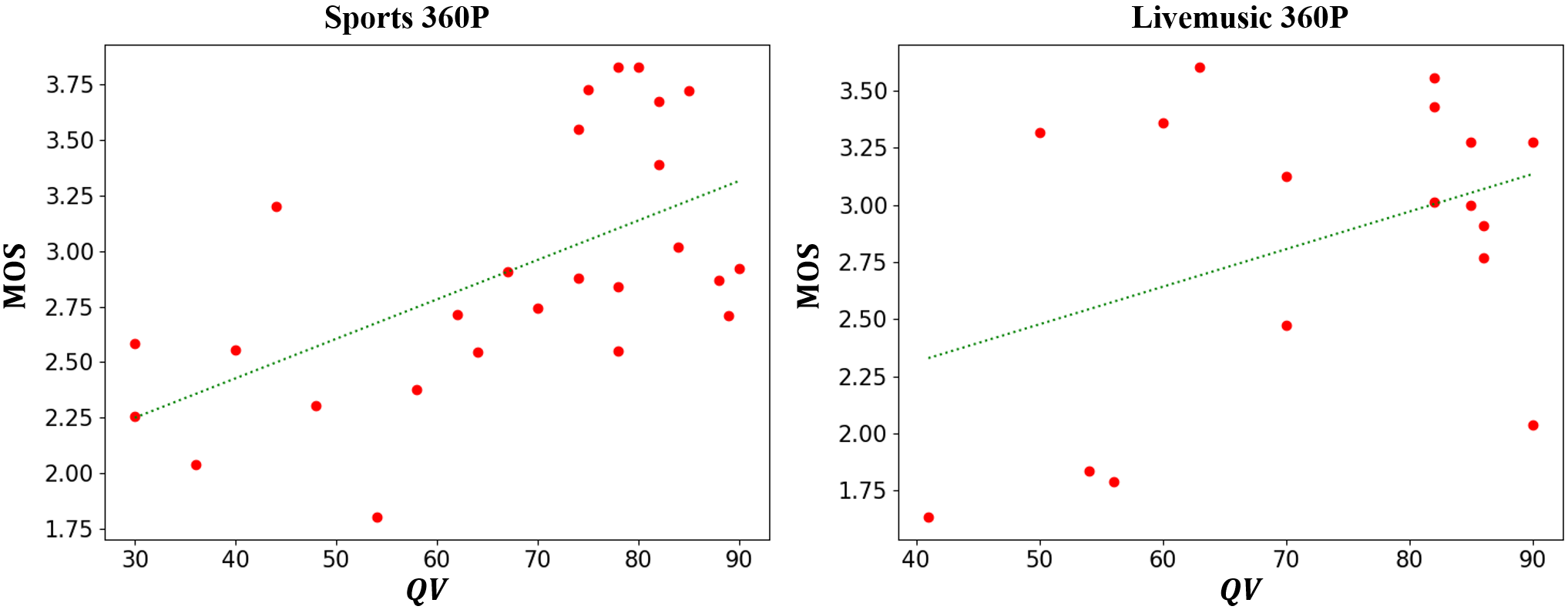}
    \caption{Scatter plot of MOS and saturation quality value $QV^*$. Each point represent a video clip from Youtube UGC.}
    \label{fig:mos_vs_qp}
\end{figure}

YouTube UGC is a large scale dataset sampled from Youtube videos. Each video clip in YouTube UGC   is accompanied by a mean-opinion-score (MOS) that provides a subjective measure of its quality. Videos are also  annotated with $15$ different content type categories. 
The dataset provides users with two versions of original videos: RAW YUV and almost lossless compressed videos using H264 CRF 10. We use the H264 CRF 10 versions. For each  clip  we sample $10$ frames, starting from the $15$th frame, and sampling every $30$ frames.   The denoised references are computed using the Python Scikit-Image \cite{van2014scikit} implementation of  the \emph{BayesShrink} wavelet denoiser \cite{chang2000adaptive}. We encode each frame with JPEG (using the Pillow Python implementation \cite{clark2015pillow}), using   $20$ different quality values ($QV_n$, $n=1,\cdots, 20$).
The  $QV_n$ ranges from $14$ (worst) to $90$ (best) with interval $4$. Our goal is to find  a saturation quality value $QV^*$, so that if we chose a $QV_n$ larger than  $QV^*$ (i.e., we increase the bitrate),  the quality of the encoded UGC has saturated. Let $\uv_{i,t}$ and $\zv_{i,t}$ denote the $i$th blocks of the $t$th frames of the UGC,   and denoised UGC signals, respectively, while $\hat{\uv}_{i,t,n}$ is the $i$th block, of the $t$th frame of the UGC encoded using $QV_n$.  
Applying the saturation criterion \eqref{eq_empirical_noise_encoding} to each block, we compute
% \begin{equation}
%     \delta_{t,n,i} = \mathbbm{1}_ {\Vert \hat{\uv}_{i,t,n} - \uv_{i,t} \Vert \leq    \Vert \hat{\uv}_{i,t,n} -  \zv_{i,t} \Vert} (\hat{\uv}_{i,t,n}),
% \end{equation}
\begin{equation}\label{eq_saturation_indicator}
 \delta_{t,n,i} = \left\{ \begin{array}{cc}
		1  & \mbox{if } \Vert \hat{\uv}_{i,t,n} - \uv_{i,t} \Vert \leq    \Vert \hat{\uv}_{i,t,n} -  \zv_{i,t} \Vert \\
		0 & \mbox{otherwise }  
	\end{array} \right. .
\end{equation}
%We compute the saturation $QV^*_{i,t}$ for each block using the saturation criteria \eqref{eq_empirical_noise_encoding}, thus
We say that the $i$th block of the $t$th frame has saturated  at quality value $QV_n$   if  $\delta_{t,n,i} = 1$. 
We define the saturation quality value $QV_{i,t}^* = QV_{n^*}$ for the $i$th block in  frame $t$ as the smallest quality value that satisfies,  for all $n \geq n^*$
\begin{equation}
    QV_{i,t}^* = QV_{n^*} \leq QV_n, \textnormal{ and } \delta_{t,n,i} = 1.
\end{equation}
If $\delta_{t,n,i}=0$ for all $n$, then we say that the block does not saturate and $QV_{i,t}^{*} = \max_{n} QV_n$. %The condition $\delta_{t,n,i} = 1$ ensures that with quality value $QV_n$, block $\hat{\uv}_{i,t}$ has saturated. 
The saturation quality value of the frame $t$ is computed as $QV_t^{*} = \textnormal{median}( \{ QV_{i,t}^{*} \}_i)$. The saturation quality value of the clip is computed as $QV^* = \textnormal{median}( \{ QV_t^{*} \}_t)$.
Video clips from the Sport and Livemusic categories with resolution 360P are used to show the correlation between MOS and saturation $QV^*$, where the MOS value used is measured for the first 10 seconds rather than the whole video. In  \autoref{fig:mos_vs_qp}, we observe positive correlation between MOS and $QV^*$. Note that a perceptual metric such as MOS depends on multiple factors, including the content quality, and not just the compression quality of the UGC content. Thus, while we do observe positive correlation it is not surprising that correlation is not perfect. 
%%%%%%%%%%%%%%%%%%%%%%%%%%%%
%%%%%%%%%%%%%%%%%%%%%%%%%%%%
%
%
%
%
\section{Conclusion}
\label{sec_conclusion}
We have formulated the problem of compression of user generated content (UGC), as   compression of a noisy/distorted source. Using classic results from rate-distortion theory, we showed that optimal UGC compression can be obtained by optimal denoising/restoration followed by optimal compression of a noiseless signal. Since in practical systems, it may be undesirable and challenging to find good denoising/restoration algorithms, we propose instead, using a denoised reference to compute distortion, and guide (regularize) the compression process, to avoid spending bitrate in encoding noise and undesirable artifacts. We perform  experiments on synthetic UGC images, and show that distortion-rate curves with denoised UGC as a reference, shares similar saturation properties as the distortion-rate curve that uses the pristine (unknown) signal as reference. We then propose a simple method to detect distortion saturation of YouTube UGC videos, and demonstrate that the Quality Parameter of a JPEG encoder, at which the distortion saturates, is positively correlated with the mean-opinion-score. 

\pagebreak
\bibliographystyle{IEEEbib}
\bibliography{refs}
%\pagebreak

%%%%

\end{document}